\documentclass[reprint,twocolumn,aps,showpacs,prb,superscriptaddress]{revtex4-2}
\usepackage{amsmath,amssymb,graphics,epsfig,epstopdf,color,multirow,array,verbatim,ulem,braket,tabularx,graphicx,color}
\usepackage[colorlinks,linkcolor=blue,citecolor=blue,urlcolor=blue]{hyperref}
\setcitestyle{square,numbers,sort&compress}
\begin{document}
\title{Interfacial superconductivity in the type-III heterostructure SnSe$_2$/PtTe$_2$}
\author{Jun Fan}
\affiliation{School of Science, Jiangsu University of Science and Technology, Zhenjiang 212100, China}
\author{Xiao-Le Qiu}
\affiliation{School of Physics and Electronic Information, Weifang University, Weifang 261061, China}
\author{Zhong-Yi Lu}
\affiliation{School of Physics and Beijing Key Laboratory of Opto-electronic Functional Materials $\&$ Micro-nano Devices, Renmin University of China, Beijing 100872,China}
\author{Kai Liu}\email{kliu@ruc.edu.cn}
\affiliation{School of Physics and Beijing Key Laboratory of Opto-electronic Functional Materials $\&$ Micro-nano Devices, Renmin University of China, Beijing 100872,China}
\author{Ben-Chao Gong}\email{bcgong@just.edu.cn}
\affiliation{School of Science, Jiangsu University of Science and Technology, Zhenjiang 212100, China}

\begin{abstract}
Interfacial superconductivity (IS) has been a topic of intense interest in the condensed matter physics, due to its unique properties and exotic photoelectrical performance. However, there are few reports about IS systems consisting of two insulators. Here, motivated by the emergence of an insulator-metal transition in the type-III heterostructure and the superconductivity in the some "special" two-dimensional (2D) semiconductors via the electron doping, we predict that 2D heterostructure SnSe$_2$/PtTe$_2$ is a model system for realizing the IS by using first-principles calculations. Our results show that due to the slight but crucial interlayer charge transfer, SnSe$_2$/PtTe$_2$ turns to be a type-III heterostructure with metallic properties and shows a superconducting transition with the critical temperature ($T_\text{c}$) of 3.73 K. Similar to the enhance electron-phonon coupling (EPC) in the electron doped SnSe$_2$ monolayer, the IS in the heterostructure SnSe$_2$/PtTe$_2$ mainly originates from the metallized SnSe$_2$ layer. Furthermore, we find that the superconductivity is sensitive to the tensile lattice strain, forming a dome-shaped superconducting phase diagram. Remarkably, at the 7\% tensile strain, the superconducting $T_\text{c}$ can increase more than twofold (8.80 K), resulting from the softened acoustic phonon at the M point and the enhanced EPC strength. Our study provides a concrete example for realizing IS in the type-III heterosturcture, which waits for future experimental verification.
\end{abstract}

\date{\today}
\maketitle

\section{INTRODUCTION}
Interfacial superconductivity (IS) emerging at the interface between two different nonsuperconducting materials has been a long-term research focus in the condensed matter physics and materials science communities \cite{Introduction1,La2CuO4,Introduction3,Introduction4}. Different from their bulk matrix materials, the two-dimensional (2D) interface can exhibit plentiful physical phenomena, including quantum metallic states \cite{QuantumMetallicStates}, topological superconductivity \cite{TSC1}, anisotropic critical magnetic fields \cite{AnisotropicFields}, Ising superconductivity \cite{ISSC1,ISSC2}, \textit{etc}. Among various IS systems, the superconductivity can be observed not only at the interface between the metal and insulator like  La$_{2}$CuO$_{4}$/La$_{1. 56}$Sr$_{0. 44}$CuO$_{4}$ \cite{La2CuO4}, but also at the interface between two band-insulators, such as LaAlO$_{3}$/SrTiO$_{3}$ \cite{Introduction1}, KTaO$_{3}$/LaAlO$_{3}$ \cite{KTaO}, EuO/KTaO$_3$ \cite{KTaO}, \textit{etc}, which subverts the understanding of the contradiction between superconductivity and insulativity. However, the oxide IS materials \cite{La2CuO4,Introduction1,KTaO} have the complex interfaces formed under the condition of the inversion symmetry and spatial translation symmetry breaking, leading to be a great challenges to study and tailor the IS properties. In addition, all the previous reports on IS are relatively infrequent and highly concentrate on experimental detections. Therefore, it is desirable to search the new IS stoichiometric materials system with a clean controllable interface.
\\
\indent 2D van der Waals (vdW) semiconductors have weak interlayer interactions and smooth interface, which avoid dangling bonds, extended defects, as well as trap states \cite{WeakInterlayerInteractions}. Thus, it is an ideal platform to construct plentiful artificial heterostructures providing the excellent properties including IS, which do not exist in one of single constituting materials. On one hand, interface effects can enhance  superconductivity as reported by the monolayer FeSe films grown on the SrTiO$_3$ substrate \cite{FeSeSrTiO3}. On the other hand, through interface engineering, the 2D vdW insulator can induce superconductivity. A paradigm example is the SnSe$_2$/graphene vdW heterostructure (vdWH), in which 2D electron gas is confined to the interface \cite{SnSe2graphene}. However, the graphene is a Dirac semimetal \cite{GrapheneDiracsemimetal1} with the high carrier mobility \cite{Graphenecarrier} rather than an insulator. Thus, one expects whether it is possible to realize superconductivity in a 2D vdWH constituting of two insulators.
\\
 \indent Naturally, the first question is to make the system metallic via constructing the heterostructure. According to the different band alignments between the two constituting insulators, heterostructures can be classified into three types \cite{3Types} (Fig. S1). Apart from the insulating properties of the type-I and type-II heterostructures which have an overlap between the band gaps of two materials (Fig. S1 (b) and S1 (c)), the conduction band minimum (CBM) of a compound is lower than the valence band maximum (VBM) of the other compound in the type-III heterostructure (Fig. S1 (d)). As a result, the type-III heterostructure has been metallized by the interlayer charge transfer, which requires a large work function difference between the two layers. Besides, with the development of 2D vdWHs computational databases \cite{GaSeSnX2, 2Dheterostructures_2}, it provides an exciting opportunity for searching type-III heterostructure candidates. The other key issue is to further manipulate the superconductivity in the metallized compounds. It is worth noting that the superconductivity can be achieved in some "special" 2D semiconductors via the electron doping or the application of external electric field, such as black phosphorus \cite{blackphosphorus}, ZrNCl \cite{ZrNCl}, MoS$_2$ \cite{MoS2}, Bi$_2$Se$_3$ \cite{Bi2Se3}, SnSe$_2$ \cite{SnSe2_1,SnSe2_2}, \textit{etc}.  Combining with the metallicity of type-III heterostructure, it is possible to realize IS by selecting the proper constituting vdW materials.
\\
\indent In this paper, we have investigated the electronic structures and superconducting properties of the heterostructure SnSe$_2$/PtTe$_2$ as a model system for the IS by utilizing the first-principles calculations. We find that SnSe$_2$/PtTe$_2$ is a type-III heterostructure with a superconducting $T_\text{c}$ of 3.73 K, which originates from the the slight but crucial interlayer charge transfer, distinguishing from the semiconductor properties of the each isolated monolayer. Combining with the superconductivity of the electron-doped SnSe$_2$ monolayer, we demonstrate that heterostructure SnSe$_2$/PtTe$_2$ is a BCS interfacial superconductor. The superconductivity can be further improved by the tensile strain. At the 7\% tensile strain, the superconducting $T_\text{c}$ can increase up to 8.80 K, resulting from the softened acoustic branch at the M point and the enhanced EPC strength. Our theoretical prediction on this new type of IS system calls for experimental validation.

\section{COMPUTATIONAL DETAILS}
\label{COMPUTATIONAL DETAILS}
\indent First-principles calculations on SnSe$_2$ and PtTe$_2$ monolayer, as well as their heterostructure are mainly performed based on density-functional theory (DFT) \cite{DFT_1, DFT_2} by employing the Quantum ESPRESSO\cite{QE_1, QE_2, QE_3} and DS-PAW \cite{DSPAW,HZW} packages. The local density approximation (LDA) \cite{LDA} of Perdew-Zunger type was adopted for the exchange-correlation functional. To examine the band alignment of SnSe$_2$/PtTe$_2$ heterostructure, the PBE-GGA functional \cite{PBE}, the HSE hybrid functional \cite{HSE}, and spin orbital coupling (SOC) effect were also considered.  The kinetic energy cutoff of the plane wave basis and the charge density cutoff were set to 80 Ry and 560 Ry, respectively. The interlayer vdW interation was described by the DFT-D3 \cite{VDW} method of Grimme \textit{et al}. To avoid the interaction of periodic images, the thickness of vacuum layer for all slabs were greater than 15 $\text{\AA}$. The Gaussian smearing method with a width of 0.0037 Ry (0.05 eV) was used for the Fermi surface broadening.  All structures were fully relaxed until the forces on all atoms were smaller than 1 $\times$ 10$^{-8}$ Ry/bohr. A $32 \times 32 \times 1$ $\textbf{k}$-point grid was adopted for the Brillouin zone sampling. The $ab$-$initio$ molecular dynamics (AIMD) \cite{AIMD} simulations were carried out at 300 K with a time step of 2 fs by using DS-PAW \cite{DSPAW,HZW}. The dynamical matrices and EPC were calculated using the density functional perturbation theory (DFPT) \cite{DFPT_1,DFPT_2} in the linear response regime. A $8 \times 8 \times 1$ $\textbf{q}$-point, one sparse $16 \times 16 \times 1$, two dense $48 \times 48 \times 1$ and $64 \times 64 \times 1$ $\textbf{k}$-point meshes were used to investigate the phonon dispersion and the EPC. The superconducting gaps is calculated by using the electron-phonon Wannier (EPW) software \cite{epw1,epw2}. The superconducting transition temperature $T_\text{c}$ is obtained via the McMillan-Allen-Dynes formula \cite{McMillan-Allen-Dynes}:
\begin{equation}
T_\text{c}=\frac{\omega_{\text{log}}}{1. 2}\text{exp} \left[ \frac{-1. 04\left( 1+\lambda \right) }{\lambda - \mu^*\left(1+0. 62 \lambda \right)} \right]
\end{equation}
where, $\mu^*$ is the effective screen Coulomb repulsion constant that was set to an empirical value of 0.10 \cite{Coulombscreen1,Coulombscreen2}, $\lambda$ is the electron-phonon coupling constant, and $\omega_{\text{log}}$ is the logarithmic average frequency of the phonons, which is given by the following formula:
\begin{equation}
\omega_{\text{log}}=\text{exp}\left[\frac{2}{\lambda}\int\frac{d\omega}{\omega}\alpha^2 F\left(\omega\right)\text{ln}\left(\omega\right)\right]
\end{equation}
\\

\begin{figure*}[htbp]
\centering
\includegraphics[width=1.9\columnwidth]{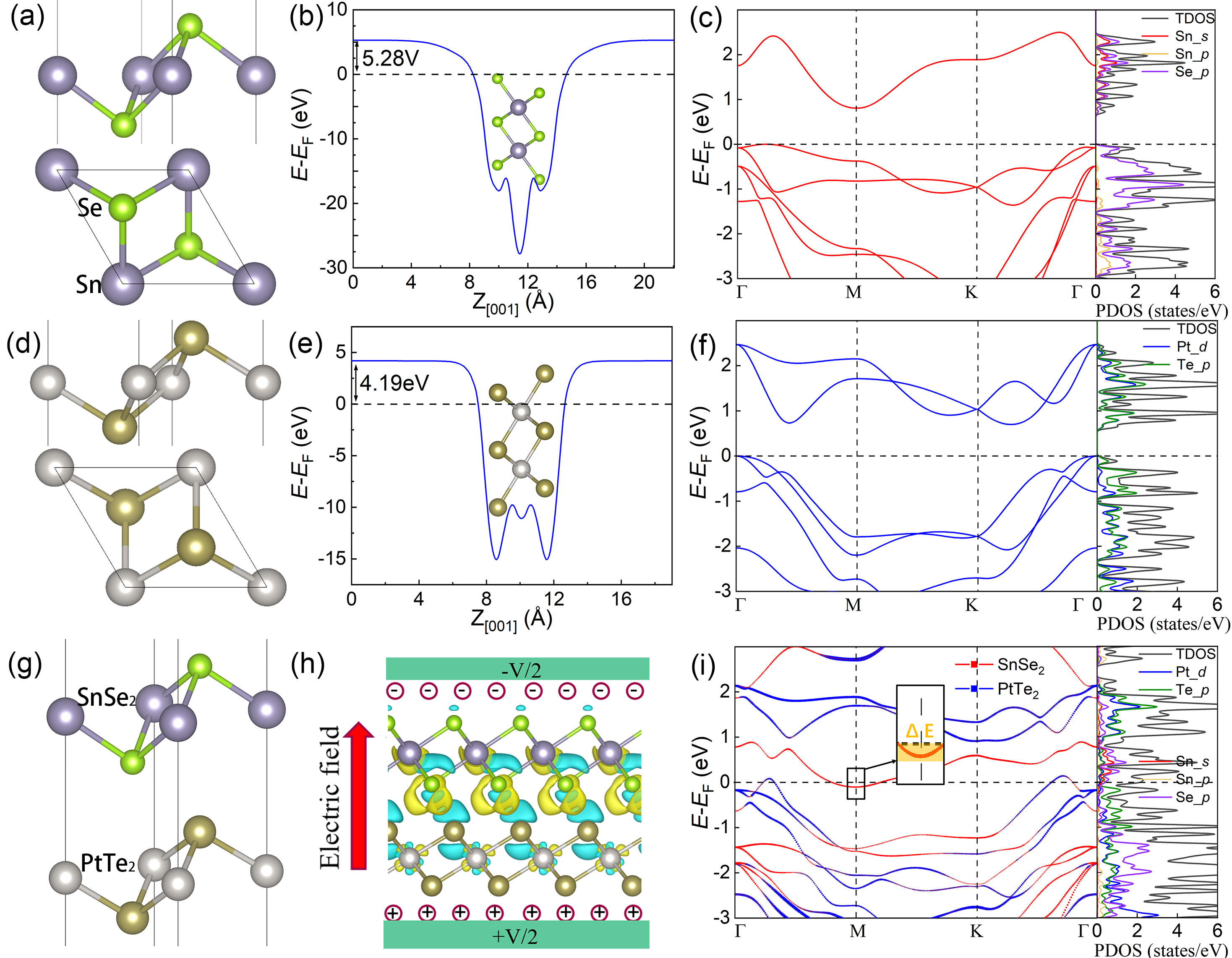}
\label{fig.1}
\caption{The atomic structure of (a) SnSe$_2$ monolayer, (d) PtTe$_2$ monolayer and (g) SnSe$_2$/PtTe$_2$ vdWH. The green, dark blue, khaki and light gray spheres represent Se, Sn, Te and Pt atoms, respectively. The work function (WF) of (b) SnSe$_2$ monolayer and (e) PtTe$_2$ monolayer. (h) Differential charge densities of the SnSe$_2$/PtTe$_2$ vdWH, as well as schematic diagram of the built-in electric field. The yellow and cyan isosurfaces show the electron accumulation and depletion areas, respectively. The isosurface of charge densities is set to 0.001 $\text{e}/\text{\AA}^3$. The band structure and PDOS of (c) SnSe$_2$ monolayer, (f) PtTe$_2$ monolayer and (i) SnSe$_2$/PtTe$_2$ vdWH. Here, the red and blue lines denote the bands weights of SnSe$_2$ and PtTe$_2$, respectively. The inset shows the electron pocket depth $\Delta E$ to evaluate the interlayer charge transfer.}
\end{figure*}

\section{RESULTS AND ANALYSIS}
\label{sec:Results AND ANALYSIS}
\indent To validate our concept, we selected the semiconductors SnSe$_2$ and PtTe$_2$ as the target materials. We first investigated the structure and electronic properties of pristine SnSe$_2$ monolayer. As shown in Fig. \hyperref[fig.1]{1(a)}, SnSe$_2$ has a sandwich-like structure with C$_{3v}$ point group. The relaxed lattice constant \textbf{a} is 3.77 $\text{\AA}$ (Table \hyperref[Table.1]{I}), consistent with the previous works \cite{latticeSnSe2_1,latticeSnSe2_2}. It has a high work function (5.28 eV) (Fig. \hyperref[fig.1]{1(b)} and Table \hyperref[Table.1]{I}), suggesting that it easily gains electrons when forming a heterostructure. To reveal the feasible exfoliation down to the monolayer, the cleavage energies (E$_{cl}$) are calculated by increasing the separation distance between the surface layer and a residual 4-layer slab. The small E$_{cl}$ (0.26 $\text{J}/{\text{m}^2}$) indicates the weak interlayer coupling (Fig. S2(a)). Figure \hyperref[fig.1]{1(c)} presents band structures and partial density of states (PDOS) of the SnSe$_2$ monolayer, in which a semiconductor behavior with a bandgap of 0.62 eV can be observed. The VBM is mainly dominated by the Se-$p$ orbitals, while the CBM is jointly contributed by the Se-$p$ and Sn-$s$ orbitals. As for the PtTe$_2$ monolayer, it has the similar structure with a lattice constant \textbf{a} of 3.94 $\text{\AA}$ (Table \hyperref[Table.1]{I}), which is similar to the previous work \cite{latticePtTe2}. The VBM is mainly derived from the Te-$p$ orbitals, while the CBM is contributed by Te-$p$ and Pt-$d$ orbitals, leaving a band gap of 0.66 eV (Fig. \hyperref[fig.1]{1(f)} and Table \hyperref[Table.1]{I}), close to the experimental value \cite{WFPtTe2}. The work function of PtTe$_2$ is 4.19 eV (Fig. \hyperref[fig.1]{1(e)}), indicating its low electronegativity. Besides, the E$_{cl}$ is calculated to be 0.66 $\text{J}/{\text{m}^2}$ (Fig. S2(a)), much higher than that (0.38 $\text{J}/{\text{m}^2}$) of the graphene \cite{CEgraphene}. We also compared the energies of bilayer with double that of monolayer $\Delta E = 2E_{\text{monolayer}}-E_{\text{bilayer}}$.
The calculated $\Delta E_{\text{PtTe}_2}$ (0.66 $\text{J}/{\text{m}^2}$) is twice as large as that of $\Delta E_{\text{SnSe}_2}$   (0.28 $\text{J}/{\text{m}^2}$). Those all imply that PtTe$_2$ has the strong interlayer coupling, in accord with the experiments \cite{latticePtTe2}. If the SnSe$_2$/PtTe$_2$ vdWH is formed, the electrons would transfer from PtTe$_2$ to SnSe$_2$. Considering the small lattice mismatch of 3.44$\%$ (Table \hyperref[Table.1]{1}), less than 5$\%$, these results make them suitable to form a type-III vdWH.
\\
\indent Then, we constructed six typical stacking configurations of SnSe$_2$/PtTe$_2$ vdWHs (Fig. S3). Noticeably, the most favorable configuration is the AA stacking (Fig. \hyperref[fig.1]{1(g)}), energetically much lower (at least 56.9 meV/f.u.) than the others (Table S1). It illustrates the interlayer interaction in SnSe$_2$/PtTe$_2$ vdWHs is strong compared with the most vdW materials \cite{StackingConfigurations_1,StackingConfigurations_2,StackingConfigurations_3}, which can be attributed to the large work function difference between the two layers. To estimate the stability of the SnSe$_2$/PtTe$_2$ vdWH, we calculated its binding energy E$_b$ by the formula
\begin{equation}
    E_b = E_{\text{SnSe}_2/\text{PtTe}_2}-E_{\text{PtTe}_2}-E_{\text{SnSe}_2}
\end{equation}
Here, $E_{\text{SnSe}_2/\text{PtTe}_2}$ is the total energy of the SnSe$_2$/PtTe$_2$ vdWH, and $E_{\text{SnSe}_2}$ and $E_{\text{PtTe}_2}$ are the energies of the isolated monolayers of SnSe$_2$ and PtTe$_2$, respectively. The calculated binding energy of SnSe$_2$/PtTe$_2$ is -0.42 eV (Table \hyperref[Table.1]{I}), and the negative value preliminarily indicates the stability of the material. To further confirm the thermal stability of this material, we perform AIMD calculations. Fig. S2(b) shows the variation of the total energy of SnSe$_2$/PtTe$_2$ over time. The results indicate that after 6 ps at T = 300 K,  the structure remains well-maintained, confirming the thermal stability of SnSe$_2$/PtTe$_2$ at room temperature. \\

\begin{table*}[htbp]
\label{Table.1}
\caption{The relaxed lattice constants, work function (WF), band gaps (${E \text{g}}$), cleavage energy ($E_{\text{cl}}$) of the SnSe$_2$ monolayer, PtTe$_2$ monolayer and SnSe$_2$/PtTe$_2$ vdWH, as well as the lattice mismatch and the binding energies ($E_{\text b}$) of the SnSe$_2$/PtTe$_2$ vdWH.}
\centering
\tabcolsep=0.015\linewidth
\begin{tabular}{ccccccc}
\hline
   & $\ \ $$a$ ($\text{\AA}$) $\ \ $& WF (eV) & $E_{\text g}$  (eV) & $E_{\text{cl}}$ ($\text{J}/{\text{m}^2}$) & Mismatch ($\%$) & $E_{\text b}$ (eV) \\
\hline
SnSe$_2$            &  3.77   &    5.28   &    0.62    &    0.26    &   -     &     -            \\
PtTe$_2$            &  3.94   &    4.19   &    0.66    &    0.66    &   -     &     -            \\
SnSe$_2$/PtTe$_2$   &  3.97   &     -     &    0.00    &     -      &  3.44   &     -0.42       \\
\hline
\end{tabular}
\end{table*}

\begin{figure}[tbp]
\centering
\includegraphics[width=0.97\columnwidth]{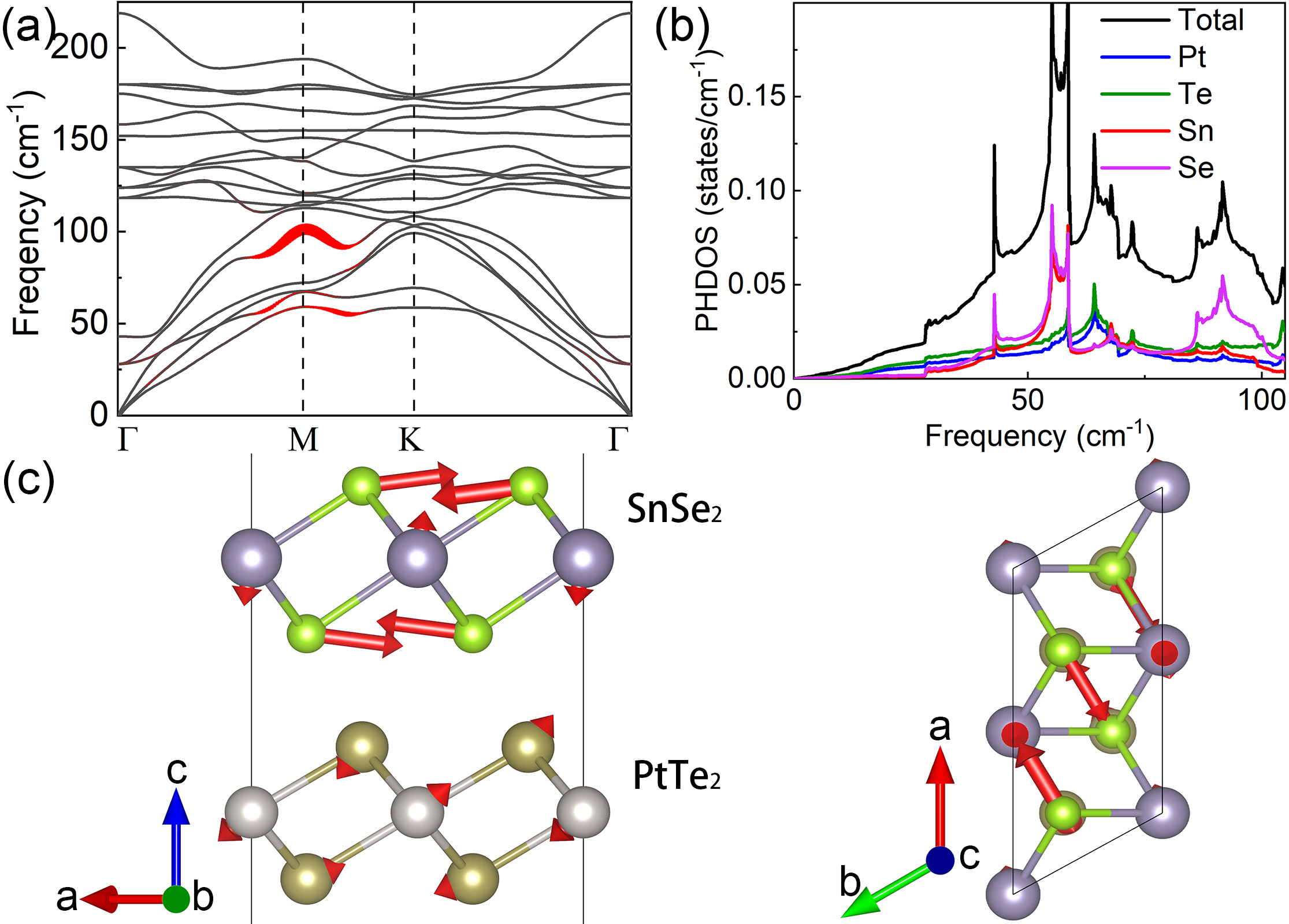}
\label{fig.2}
\caption{(a) Phonon dispersion spectrum of the SnSe$_2$/PtTe$_2$ vdWH, where the size of the red dots is proportional to the phonon linewidth. (b) Phonon density of states (PHDOS) of the SnSe$_2$/PtTe$_2$ vdWH. (c) Side and top views of phonon vibration modes with the frequency of 101 cm$^{-1}$ around the M point. The length of the arrows represent the amplitude of the atomic vibrations.}
\end{figure}
\indent We next studied the electronic properties of the SnSe$_2$/PtTe$_2$ vdWH. As shown in (Fig. \hyperref[fig.1]{1(i)}), there are two bands crossing the Fermi level, indicating the metallic behavior of the SnSe$_2$/PtTe$_2$ vdWH. One conduction band mainly originates from the SnSe$_2$ layer, while the other valence band is contributed by the hybridized orbitals of the two layers. And it shows an electron pocket at the M point and a hole pocket near the G point, demonstrating a clear characteristic of type-III heterostructure formed by two semiconductors. This vdWH would exist of the interlayer charge transfer. We further calculated differential charge density
\begin{equation}
   \Delta \rho = \rho_{\text{SnSe}_2/\text{PtTe}_2}-\rho_{\text{PtTe}_2}-\rho_{\text{SnSe}_2}
\end{equation}
Here, $\rho_{\text{SnSe}_2/\text{PtTe}_2}$ is the charge density of SnSe$_2$/PtTe$_2$, while $\rho_{\text{PtTe}_2}$ and $\rho_{\text{SnSe}_2}$ are the charge densities of the isolated SnSe$_2$ and PtTe$_2$ monolayers, respectively. Obviously, SnSe$_2$ layer accumulates the negative charges and PtTe$_2$ layer accumulates the positive charges, resulting in a built-in electric field from the PtTe$_2$ layer to the SnSe$_2$ layer (Fig. \hyperref[fig.1]{1(h)}). In other words, the electrons transfer from the VBM of PtTe$_2$ layer to the CBM of SnSe$_2$ layer, showing as the edges of CBM (VBM) dropping below (rising above) the Fermi level, respectively (Fig. \hyperref[fig.1]{1(i)}).  Besides, we also check the band structure of SnSe$_2$/PtTe$_2$ vdWH with the HSE (Fig. S4 ) and with the inclusion of the SOC effect (Fig. S5).  As we can see, these factors do not change the characteristic of type-III heterostructure.
\\
\begin{figure}[tbp]
\centering
\includegraphics[width=0.97\columnwidth]{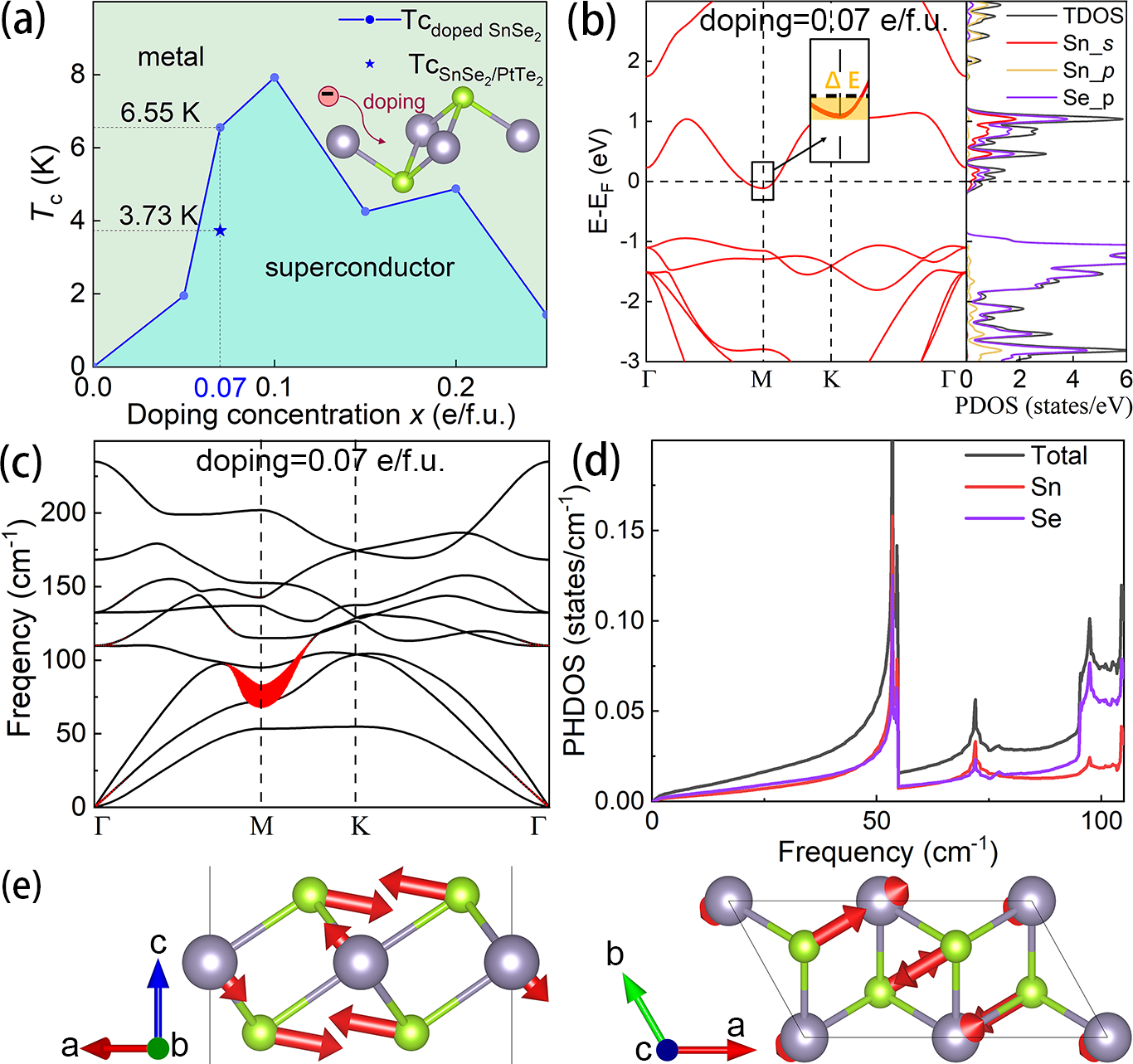}
\label{fig.3}
\caption{(a) The superconducting $T_\text{c}$ of monolayer SnSe$_2$ as a function of electron doping concentration x, as well as the value of SnSe$_2$/PtTe$_2$ vdWH (blue star). Here, the lattice constant $a$ is fixed at 3.97 $\text{\AA}$. (b) Band structure and PDOS, (c) Phonon dispersion spectrum with the momentum $q$-resolved EPC strength ($\lambda$$_q$) and (d) PHDOS of the SnSe$_2$ monolayer at a doping concentration of 0.07 e/f.u.. (e) Side and top views of phonon vibration modes with the frequency of 75 cm$^{-1}$ around the M point.}
\end{figure}
\indent Since SnSe$_2$ can become a superconductor by the electron doping \cite{SnSe2_1,DopSnSe2,supplement1}, the metallic SnSe$_2$/PtTe$_2$ vdWH motivated us to explore the possible superconductivity. The phonon spectra with the momentum ($q$)-resolved EPC strength ($\lambda$$_q$) of the SnSe$_2$/PtTe$_2$ vdWH is shown in Fig. \hyperref[fig.2]{2(a)}. The non-negative vibrational frequencies throughout the Brillouin zone indicates that SnSe$_2$/PtTe$_2$ vdWH has dynamical stability. From the calculated phonon linewidths marked by the size of red dots, the EPC strength is especially high near the M point. The largest contribution to the EPC comes from the optical branch in the frequency range of 85-101 cm$^{-1}$. Meanwhile, the acoustic branches in the frequency range of 55-70 cm$^{-1}$ also contribute significantly to the EPC. Most of them stem from the vibrations of Se atoms as demonstrated by the the phonon density of states (PHDOS)(Fig. \hyperref[fig.2]{2(b)}). To obtain the vibrational patterns of the branch with the largest EPC at the M point, a $2 \times 1 \times 1$ supercell was adopted. Thus, the M point of the unit-cell folds to the $\Gamma$ point of the supercell. It can be seen that opposite vibrations of Se atoms in the $ab$ plane play a dominant role, while the vibrations of other atoms are negligible (Fig. \hyperref[fig.2]{2(c)}). This reflects that the superconductivity in SnSe$_2$/PtTe$_2$ vdWH may originated the conductive SnSe$_2$ layer. Based on the McMillian-Allen-Dynes formula, the calculated superconducting $T_\text{c}$ of SnSe$_2$/PtTe$_2$ vdWH converges to 3.74 K, close to the measured value of the SnSe$_2$/graphene (4.84 K) \cite{SnSe2graphene}, ultrathin Li-intercalated SnSe$_2$ (4.8 K) \cite{SnSe2_1} and ionic-liquid gated SnSe$_2$ (3.9 K) \cite{ionic-liquidGatedSnSe2}.
\\
\indent To further investigate the mechanism of superconductivity in the SnSe$_2$/PtTe$_2$ vdWH, we made a comparative study on the electron doped SnSe$_2$ monolayer. Here, the approach to simulate the charge doping concentration is by changing the total number of electrons of the system with a compensating jellium background. To rule out the influence of structural factors, the lattice constant of charge doped SnSe$_2$ monolayer is fixed at the same value of SnSe$_2$/PtTe$_2$ vdWH (a=3.97 $\text{\AA}$). Figure \hyperref[fig.3]{3(a)} shows the superconducting $T_\text{c}$ as a function of electron doping concentration $x$. Apparently, with the increasing electron doping, the superconducting $T_\text{c}$ first increases and then decreases in a dome-shape manner. The highest value can reach 7.9 K at the doping of 0.1 e/f.u.. It is highly in line with the trend of the phonon softening at the M point (Fig. S6). To accurately evaluate the doping concentration $x$ compared with the interlayer charge transfer in the SnSe$_2$/PtTe$_2$ vdWH, we define the electron pocket depth $\Delta E = E_{F} - E_{\text{CBM}}$ (insets of Fig. \hyperref[fig.3]{3(b)} and Fig. \hyperref[fig.1]{1(i)}). Here, $E_{\text{CBM}}$ is the energy of the CBM edge, and $E_F$ is the Fermi level. Then, we obtain the charge transfer from the PtTe$_2$ layer to SnSe$_2$ layer is about 0.07 e/f.u.. While, at the same doping concentration, the superconducting $T_\text{c}$ of the SnSe$_2$ monolayer is 6.55 K, higher than that of SnSe$_2$/PtTe$_2$ (3.73 K), which may be owing to the more uniform electron doping via the jellium model. Analogously, we analyzed the phonon dispersion and EPC properties of the 0.07 e/f.u. doped SnSe$_2$ monolayer, as shown in Fig. \hyperref[fig.3]{3(c)-(e)}. The phonon branch with frequency range of 75-100 cm$^{-1}$ at the M point contributes to a strong EPC (Fig. \hyperref[fig.3]{3(c)}). The corresponding vibrations mainly depend on Se atoms (Fig. \hyperref[fig.3]{3(e)}), identical to the case of SnSe$_2$/PtTe$_2$ vdWH (Fig. \hyperref[fig.2]{2(c)}). On the other hand, we also explored the PtTe$_2$ monolayer with hole doping. At a doping concentration of 0.07 h/f.u., it does not show up the superconductivity. These results suggest that it is the SnSe$_2$ layer that obtains the electrons from PtTe$_2$ layer and then induces the IS in the SnSe$_2$/PtTe$_2$ vdWH.
\\
\begin{figure}[htp]
\centering
\includegraphics[width=0.80\columnwidth]{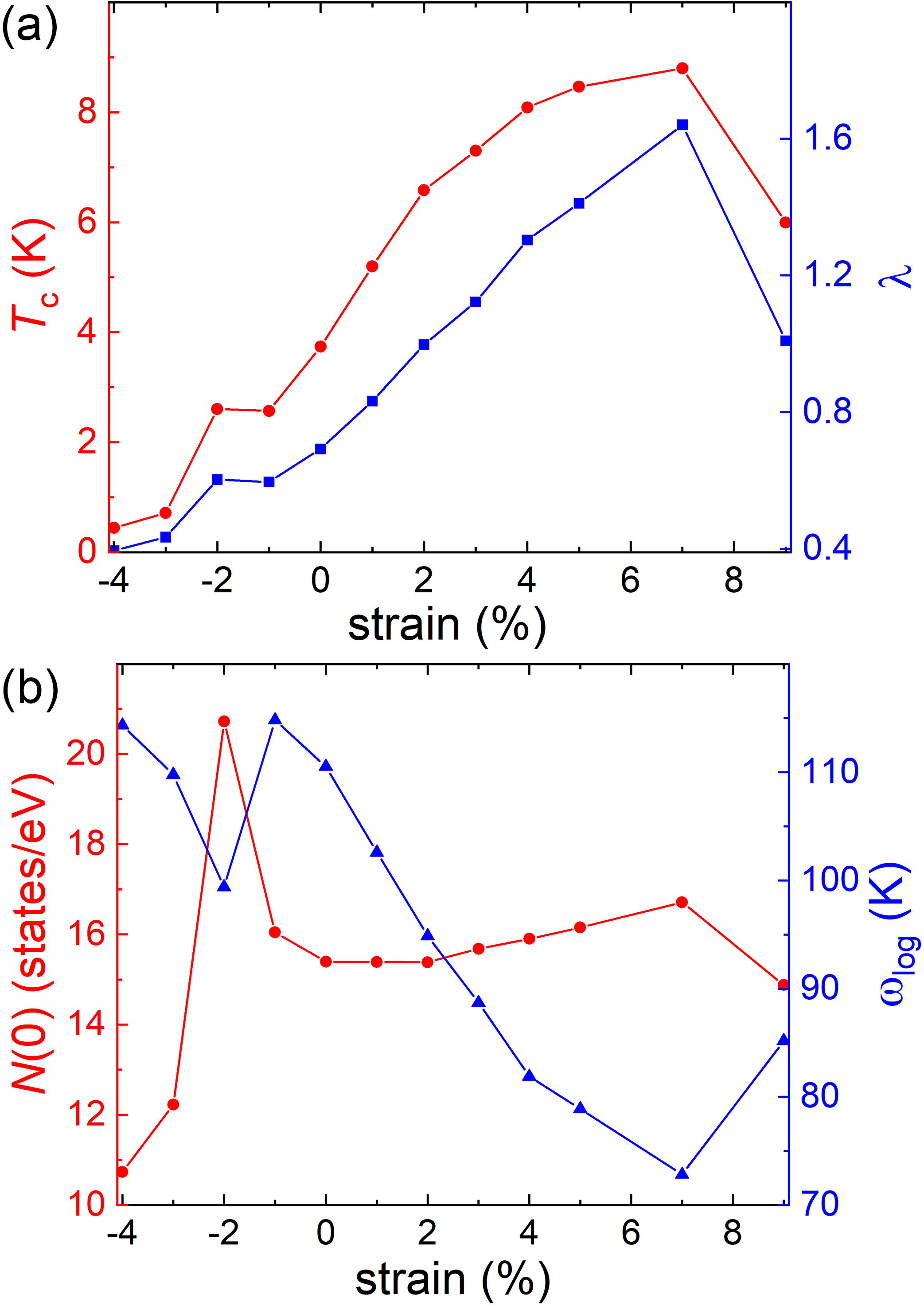}
\label{fig.4}
\caption{(a) Superconducting $T_\text{c}$ (red) and EPC constant $\lambda$ (blue) of the SnSe$_2$/PtTe$_2$ vdWH under different lattice strains. (b) Density of states at the Fermi level $N\left(0\right)$ (red) and logarithmic average frequency $\omega_{\text{log}}$ (blue) under different strains.}
\end{figure}
\indent Inspired by the fact that the $T_\text{c}$ of the electron doped SnSe$_2$ monolayer is sensitive to the lattice constant (Fig. S7), we further probed the evolutions of the superconductivity of the SnSe$_2$/PtTe$_2$ vdWH with the lattice strain. As shown in Fig. S8(b) and Fig. \hyperref[fig.4]{4(a)}, the SnSe$_2$/PtTe$_2$ vdWH maintains dynamical stability and superconducting under the strain from -4$\%$ (compressive strain) to 9$\%$ (tensile strain). Apparently, the compressive stress hampers the the superconductivity, as a whole, except at the -2$\%$ strain where the DOS at the Fermi level ($N\left(0\right)$) (Fig. \hyperref[fig.4]{4(b)}) shows a peak inducing the elevation of the EPC constant $\lambda$ and $T_\text{c}$ (Fig. \hyperref[fig.4]{4(a)}). In contrast, the tensile strain is prone to improve the $T_\text{c}$, which is explained by the softened acoustic phonon at the M point (Fig. S8). It shows a dramatic decrease of the logarithmic average frequency $\omega_{\text{log}}$ and a slight increase of $N\left(0\right)$ under the strains from 0 to 7 $\%$ (Fig. \hyperref[fig.4]{4(b)}). According to the McMillan-Allen-Dynes formula, the superconducting $T_\text{c}$ is exponentially related to $\lambda$, but is linearly dependent on $\omega_{\text{log}}$. Hence, the $\lambda$, as well as $T_\text{c}$, first increases and then decreases with increase of the tensile strain. Remarkably, at the 7\% tensile strain, the superconducting $T_\text{c}$ can increase more than twofold (8.80 K).
\\
\section{discussion and summary}
\indent Our strategy for realizing the IS has several advantages. First, it is a new approach based on the theoretical prediction, while previous IS systems highly depend on experimental measurements. Due to the enormous type-III vdWHs database \cite{2Dheterostructures_2}, it is also convenient for experimentalists to search more IS candidates. Second, the type-III vdWHs own the divinable smooth surfaces, compared with those of oxide IS systems. Last but not least, our proposal is not limited to the type-III vdWHs. In the view of the fact that the interlayer charge transfer results from the built-in electric field, the external vertical electric field can adjust the vdWHs to induce transitions among the three type configurations, demonstrated by the GaSe/SnX$_2$ (X = S, Se) \cite{GaSeSnX2}, bilayer-NiI$_2$/bilayer-In$_2$Se$_3$ \cite{bilayerNiI2_bilayerIn2Se3}, Janus SnSSe/phosphorene \cite{DopSnSe2}, \textit{etc}. Thus, IS candidates can be extended to the other two types of vdWHs via the appropriate electric field.
\\
\indent In order to facilitate experimental studies, we have simulated the scanning tunneling spectra (STS) curves  (Fig. S9) and the scanning tunneling microscopy (STM) images (Fig. S10) for different terminated surfaces of SnSe$_2$/PtTe$_2$ vdWH. The quasiparticle density of states at low temperatures (Fig. S9) show a U-shape full gap, consistent with the electron-phonon coupling mechanism. Meanwhile, as we can see, there are perfect triangle patterns on both the Se-terminated (top) and Te-terminated (bottom) surfaces in different energy windows (Fig. S10), suggesting its high quality without any obvious structural distortions. Hence, the SnSe$_2$/PtTe$_2$ vdWH would be a good model system for experimentally realizing the IS.
\\
\indent Due to the pressure-induced superconductivity in 1$T$-SnSe$_2$\cite{SnSe2-pressure,SnSe-pressure}, it deserves to compare that with the strain effect in SnSe$_2$/PtTe$_2$ vdWH.  In previous experiments, the superconductivity in pristine 1$T$-SnSe$_2$ emerges around 18.6 GPa, which brings with an insulator-metal transition, and keeps robust ($T_\text{c}$ = 6.1 K) under high pressure from 30.1 to 50.3 GPa \cite{SnSe2-pressure}. Here, we studied the evolution of superconductivity for the SnSe$_2$/PtTe$_2$ vdWH with the lattice strain from -4\% to 9\%. The compressive and tensile strains could correspond to the positive and negative pressures, respectively. But due to the lack of constrain along the c direction and the constituting materials are different for the SnSe$_2$/PtTe$_2$ vdWH, its superconducting behavior is distinct from that of bulk 1$T$-SnSe$_2$. For instance, the compressive strain on the SnSe$_2$/PtTe$_2$ vdWH suppresses the superconducting $T_\text{c}$ (Fig. \hyperref[fig.4]{4(a)}) while the positive pressure on the bulk  SnSe$_2$ is conducive to the superconductivity. Thus, SnSe$_2$/PtTe$_2$ vdWH is also a good platform for exploring the strain effect on the superconducting $T_\text{c}$ of IS systems.
\\
\indent In summary, we have theoretically built a model system SnSe$_2$/PtTe$_2$ vdWH and explored its interfacial superconducting properties by using first-principles calculations. Our results show that the monolayers SnSe$_2$ and PtTe$_2$ are semiconductors with a big difference of work function. After constructing the heterostructure, the SnSe$_2$/PtTe$_2$ vdWH turns into a metal and exhibits the superconductivity below $T_\text{c}$ $\sim$ 3.73 K. This IS mainly originates from the interlayer charge transfer, which is analogous to electron doped SnSe$_2$ monolayer. Under the tensile strain, the superconducting $T_\text{c}$ can be improved as high as 8.80 K, which results from the enhanced electron-phonon coupling and the softened acoustic phonon at the M point. Our study provides an useful guidance for experimentalists in selecting suitable IS systems.
\begin{acknowledgments}
We wish to thank Zhenfeng Ouyang, Chang-Jiang Wu and Jing Jiang for helpful discussions. This work was supported by the National Key R\&D Program of China (Grants No. 2022YFA1403103 and No. 2019YFA0308603), the National Natural Science Foundation of China (Grants No. 12304167), the Shandong Provincial Natural Science Foundation of China (Grant No. ZR2023QA020). The authors gratefully acknowledge HZWTECH and the Physical Laboratory of High Performance Computing at Renmin University of China for providing computation facilities.
\end{acknowledgments}

\bibliography{./Reference}

\end{document}